# A four dimensional map for escape from resonance: negative energy modes and nonlinear instability

Caroline G. L. Martins, P. J. Morrison and C. Curry

University of Texas at Austin, Department of Physics, Institute for Fusion Studies, Austin - TX 78712, USA

**ABSTRACT**

Positive definiteness of a Hamiltonian expanded about an equilibrium point provides only a necessary condition for stability, a criterion known as Dirichlet's theorem. The reason that this criterion is not necessary for stability is because of the possible existence of negative energy modes, which are linearly stable modes of oscillation that have negative energy. When such modes are present, the Hamiltonian is, in general, indefinite. Although such systems with negative energy modes are linearly stable (spectral stable), they are unstable to infinitesimal perturbations under the nonlinear dynamics. In the present work we study this kind of nonlinear instability with the simplest nontrivial four dimensional area-preserving map, which has a cubic degree of freedom, that was designed to mimic the behavior of a Hamiltonian system with one positive and one negative energy mode, and a quadratic degree of freedom, that allows eventual escapes in phase space, usually called as Arnold diffusion.

**PACS:**

# 1. Introduction

Questions about the stability of the solar system, steady rotating fluids and floating bodies have been strong motivations for the theoretical researches on stability theory. The technical term "stability" has its origin in mechanics, as was reported in 1749 in Euler's work [26], even though, nowadays, the interest in the stability of motion is broader and plays a crucial role in economical models, numerical algorithms, nuclear physics and control theory, the last one being applied, for example, in the fields of mechanical and electrical engineering [25].

The essence of our modern concept of stability is usually thought to be developed at the end of the nineteenth century, most precisely at 1892 by Lyapunov [28], but, as a matter of fact, it can be found in Lagrange's work from 1788 [27]. In other words, Lagrange states that an equilibrium is stable when neighboring solutions remain close to the equilibrium, which agrees with the concept of stability in the sense of Lyapunov. Moreover, Lagrange wrote the theorem where, if the system is conservative, a state corresponding to zero kinetic energy and minimum potential energy is a stable equilibrium point [27].

Magnetohydrodynamic theory (MHD) is often used to analyze fusion confinements by describing the plasma as a single magnetized fluid, and, since Lagrange's theorem assumes that the analysis of the potential energy provides a necessary and sufficient condition for stability, a famous example can be illustrated as the 'MHD energy principle', where the primary interest is to analyze, simply, the stability of a particular equilibrium, according to changes in $\partial W$ [29].

More generally, positive definiteness of a Hamiltonian expanded about an equilibrium point, provides only a necessary condition for stability, a criterion known as Dirichlet's theorem [3]. The reason that this criterion is not necessary for stability is because of the possible existence of negative energy modes (NEMs) [5]. NEMs are modes of oscillation that have negative energy, and when such modes are present, the Hamiltonian is in general indefinite [1]. It has been argued in [7, 8] that, although such systems with negative energy modes are linearly stable, they are unstable to infinitesimal perturbations under the nonlinear dynamics.

It is well known, in two dimensional systems, that invariant tori are closed curves which provide barriers to transport. If a stable fixed point is surrounded by a nested set of such curves, the equilibrium is nonlinearly stable. However, in higher dimensions, invariant tori do not provide barriers to the transport, modifying the stability of the equilibrium, and allowing an avenue for escape, which is usually called as Arnold diffusion [30].

In the present work we analyze the equilibrium stability of a simplest nontrivial four dimensional area-preserving map, which has a cubic degree of freedom, that was designed to mimic the behavior of a Hamiltonian system with one positive and one negative energy mode, and a quadratic degree of freedom, that allows eventual escapes in phase space. A special regular orbit present in the cubic degree of freedom is examined in details, when the coupling is activated, which plays an important role in the escape of chaotic trajectories.

In section II we introduce a brief review on the stability of equilibria and negative–energy modes (NEMs). In section III we present the cubic map that was designed to mimic the behavior of a Hamiltonian system with one positive and one negative energy mode. The four dimensional map generated by the coupling between the cubic and a quadratic map is presented in section IV, as well as analyses on the escape of chaotic trajectories in phase space (Arnold diffusion), and details about the special regular orbit, termed "triangular torus". Finally, in section V, we present the conclusions.

## 2. Stability of equilibria and negative–energy modes (NEM)

Models of physical systems often produce sets of differential equations of the form $\dot{z}^i = V^i(z)$ with $i = 1,...,n$, where $z = (z^1, z^2,...,z^n)$. Let's consider such dynamical system with an equilibrium point, $z_e$, satisfying $V(z_e) = 0$. The behavior of solutions near such equilibrium points is called *stability of equilibria* and this idea is formalize by the following stability proof definition [1]:

> *The equilibrium point $z_e$ is said to be stable if, for any neighborhood N of $z_e$ there exists a sub-neighborhood $S \subset N$ of $z_e$ such that if $z(t=0) \in S$, then $z(t) \in N$ for all time $t > 0$.*

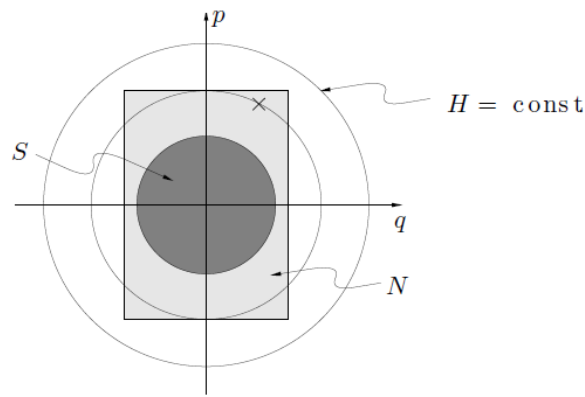

Figure 1 – Neighborhoods and sub-neighborhoods for stability of the simple harmonic oscillator

Figure 1 shows the phase space for the simple harmonic oscillator, and illustrates the regions used in the definition of stability. In this figure the circles are surfaces of constant energy. Here we have chosen as a neighborhood *N* the rectangular region in which we have marked an initial condition by the symbol "×". Since trajectories move around on the circles of constant *H*, it is clear that, in a short time, the trajectory starting at × will leave *N*, in spite of the fact that the equilibrium point at the origin is stable. However, if we choose initial conditions inside the circular sub-neighborhood *S*, which is defined as the region bounded by a $H = constant$ surface, contained in *N*, then the trajectory will remain in *N* for all time. Thus $H = constant$ surfaces serve as a handy means of defining sub-neighborhoods.

When $z(t)$ is determined from the linearized dynamics,

$$\delta \dot{z}^i = \frac{\partial V^i}{\partial z^j}(z_e) \delta z^j \tag{1}$$

where now $z(t) \approx z_e + \delta z$, and this dynamics is stable according to the above definition, we say that $z_e$ is *linearly stable*.

The stability under the full nonlinear dynamics, $V(z)$, is called *nonlinear stability*. Equilibria that are unstable under nonlinear dynamics, yet stable under linear dynamics, are said to be *nonlinearly unstable*. Last, but not least, equilibria that have all eigenvalues of its linearization (equation 1) as pure imaginary, are called *spectrally stable*. The relation between the different kinds of stabilities can be found in the following scheme,

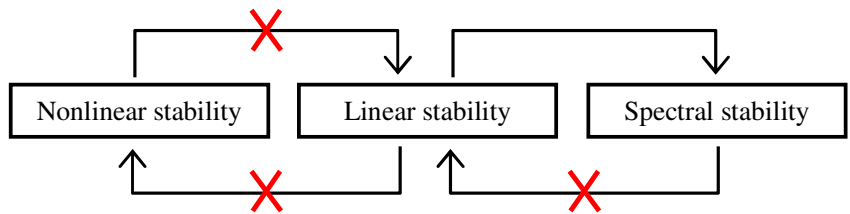

and the examples that demonstrate the relations are described in the next paragraphs.

When the Hamiltonian has a separable form, $H(q,p) = (p^2/2) + V(q)$, and old theorem due to Lagrange states that an equilibrium point with $p_e = 0$, and $q_e$ being a local minimum of $V(q)$, is stable. It is intuitive to think that the converse should be true, but a counterexample [2] shows the opposite. Consider the set of equations,

$$V(q) = \begin{cases} e^{-(1/q^2)} \cos(1/q) & q \neq 0 \\ 0 & q = 0 \end{cases} \tag{2}$$

The equilibrium position, $q_e = 0$, is stable, but due to the wild oscillation that occurs as $q_e \to 0$, the origin is not a local minimum. However, with some relatively mild restrictions on $V(q)$, *Lagrange's theorem* is both necessary and sufficient for stability for Hamiltonians of this restricted form.

One might be fooled into thinking that nonlinear stability implies linear stability; however, with a little thought you can see that this is not true. The one degree-of-freedom system with potential $V(q) = q^4/4$, has an equilibrium point $q_e = 0$, and it is clear that this is nonlinearly stable since $H(q, p)$ defines good sub-neighborhoods. This example also reveals why spectral stability does not imply linear stability [1].

Hamiltonian systems are not always of the separable form, but are instead general functions of $q$ and $p$. When this is the case another old theorem, which is sometimes called *Dirichlet's theorem*, gives a sufficient condition for stability [3]. If in the vicinity of an equilibrium point, surfaces of $H = constant$ define a family of good neighborhoods, then the equilibrium is nonlinearly stable. For "well-behaved" Hamiltonians one need only analyze the matrix $\partial^2 H(z_e)/\partial z^i \partial z^j$, where $z = (q, p)$. If this matrix is definite, i.e. it has no zero eigenvalues and all of its eigenvalues have the same sign, then we have stability. Observe that $H$ could, in fact, be an energy maximum. This can occur for rigid-body dynamics and is typically the case for a localized vortex in fluid mechanics [1].

There is an example due to Cherry that illustrates two things: that Dirichlet's theorem is not necessary and sufficient for stability and that linear stability does not imply nonlinear stability. *Cherry's Hamiltonian* is [4],

$$H(q_1, p_1, q_2, p_2) = \frac{1}{2}\omega_2(p_2^2 + q_2^2) - \frac{1}{2}\omega_1(p_1^2 + q_1^2) + \frac{1}{2}\alpha[2q_1 p_1 p_2 - q_2(q_1^2 - p_1^2)] \qquad (3)$$

where $\omega_1$ and $\omega_2$ are adjustable frequencies, where $\omega_{1,2} > 0$, and $\alpha$ is a nonlinear parameter. The minus sign at the term $-(\omega_1/2)(p_1^2 + q_1^2)$, from equation (3), makes the matrix $\partial^2 H(z_e)/\partial z^i \partial z^j$ not definite, and this term represents what is called negative-energy mode (NEM) [5]. Observe that this minus sign cannot be removed by a time-independent canonical transformation and, in the typical case, cannot be removed by any canonical transformation.

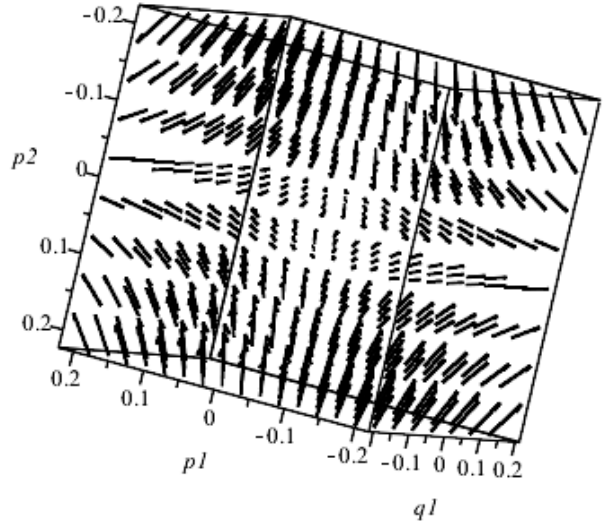

Figure 2 – Three dimensional vector field of Cherry's Hamiltonian from equation (3) at the section $q_2 = 0$, with $\omega_2 = 2\omega_1$ and $\alpha \neq 0$, emphasizing the explosive growth of the initial conditions surrounding the equilibrium point $q_1 = q_2 = p_1 = p_2 = 0$.

Figure 2 shows the three dimensional vector field of Cherry's Hamiltonian from equation (3) at the section $q_2 = 0$, with $\omega_2 = 2\omega_1$ and $\alpha \neq 0$. In spite of the linear stability of the origin, $q_1 = q_2 = p_1 = p_2 = 0$, there are analytical solutions for the current case which demonstrate that any initial condition in the neighborhood of the origin diverges in finite time, showing the nonlinear instability feature of this particular case. Such behavior is referred to as explosive growth and is characteristic of systems that possess both NEMs and resonance [1].

Explosive behavior is to be expected in systems with both positive and negative energy modes, and likely Cherry's Hamiltonian, other well-known example is the *Three-wave Hamiltonian Problem* [6]. One can find an in-depth stability analysis of the problem published in two parts at [7, 8], emphasizing "negative energy resonances" (involving two negative energy modes and one of positive energy) and "positive energy resonances" (two positive energy modes and one of negative energy).

The three-wave problem is an example of a system with three resonances that are driven unstable by cubic terms in the Hamiltonian. Discrete area-preserving maps containing cubic terms can be modeled in order to mimic the behavior of Hamiltonians that present NEMs, and the same explosive behavior discussed in the current section could appear, more details can be found in the next section.

## 3. Discrete cubic map

It is well-known that the study of discrete maps has become an important part of nonlinear systems analysis [9-11]. These maps have been of interest to model, for example, problems in biology of populations, physical and engineering problems, and even chemical problems [12-17]. There are several simple maps that often arise in practical applications, such as the family of cubic maps. In the late 70's a one dimensional cubic map was considered by R. M. May in connection with ecological phenomena [18], and a dissipative two dimensional cubic map was considered by P. Holmes in the study of strange nonlinear oscillations in a buckled beam undergoing forced lateral vibrations [19]. After that some relevant studies concerning different types of cubic maps were published due to their dynamical richness. Similarities between the bifurcation ratio of the one dimensional cubic map and the quadratic map are presented at [22] and a new kind of intermittent transition to chaos regarding a one dimensional cubic map can be seem at [23]. Works regarding dissipative cubic maps were published as well, in order to analyze, for example, jumps among independent attractors in the one dimensional cubic map, in the presence of noise [20], correlations between crises of attractors and the locations of unstable orbits [21] and comparisons between the bifurcation ratio of the dissipative two dimensional cubic map and the Hénon map [22].

We present in this work a discrete map that mimics the behavior of a Hamiltonian system with one positive and one negative energy mode. It can be compared as a modification of the dissipative two dimensional cubic map introduced by Holmes [19]. The new map is symplectic (area-preserving), and it is described by the following set of equations,

$$p_{n+1} = -q_n$$
$$q_{n+1} = p_n + q_n t - q_n^3 \qquad (4)$$

where $(q, p)$ are canonical variables of position and momentum.

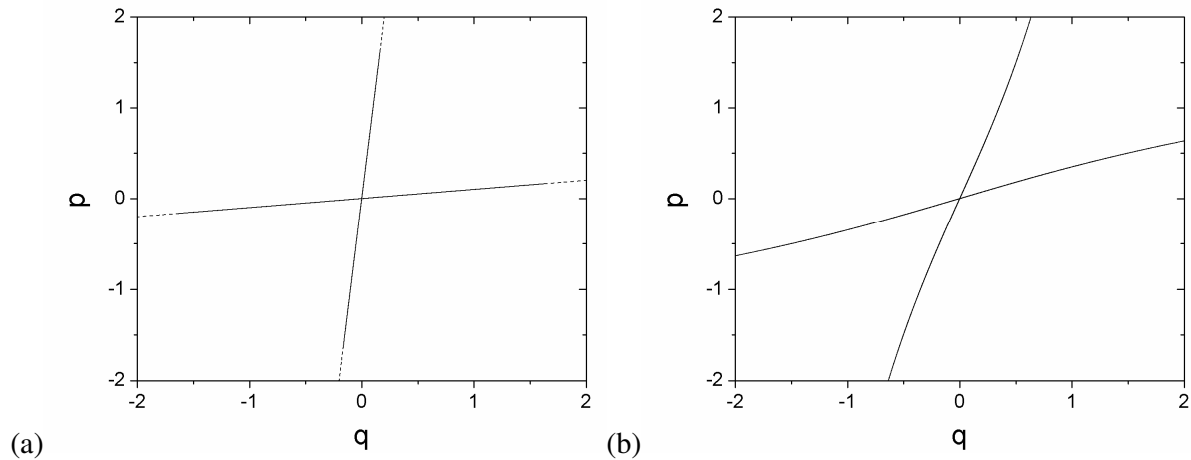

(a)  (b)

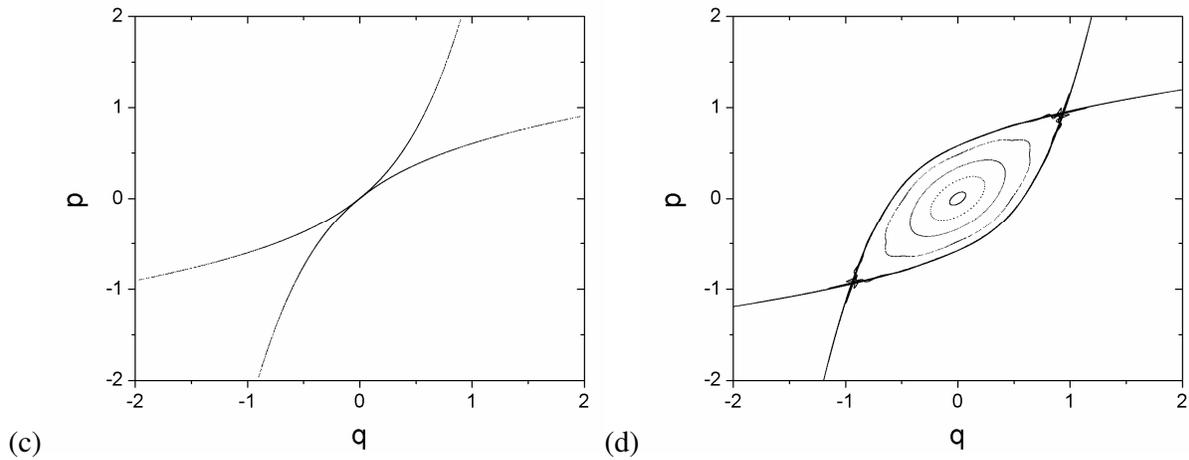

(c)  (d)

Figure 3 – Phase space of the cubic map from equation (4), for (a) $t = -10.0$, (b) $t = -3.1$, (c) $t = -2.0$ and (d) $t = -1.1$. An inverse tangent bifurcation occurs while we increase the parameter $t$, i.e. one can find in (a) a hyperbolic fixed point at $q = p = 0$, and in (d) an elliptic fixed point at $q = p = 0$, and two hyperbolic fixed points at $q = p \approx 1$ and $q = p \approx -1$.

Figure 3 presents the phase space of the cubic map from equation (4), for (a) $t = -10.0$, (b) $t = -3.1$, (c) $t = -2.0$ and (d) $t = -1.1$. One can find in 3(a) a hyperbolic fixed point at $q = p = 0$, and in 3(d) an elliptic fixed point at $q = p = 0$, and two hyperbolic fixed points at $q = p \approx 1$ and $q = p \approx -1$, emphasizing an inverse tangent bifurcation that occurs while we increase the parameter $t$. For the case showed in figure 3(a), the system presents linear stability at the origin, $q = p = 0$, but due to the explosive growth of initial condition located in the neighborhood of the origin, the system presents nonlinear instability as well. However, after the inverse tangent bifurcation, figure 3(d) shows invariant tori close enough to the elliptic fixed point in the origin, acting as sub-neighborhoods in the stability proof from page 2. In resume, the inverse tangent bifurcation turned the nonlinear unstable system of figure 3(a), into a nonlinear stable system of figure 3(d).

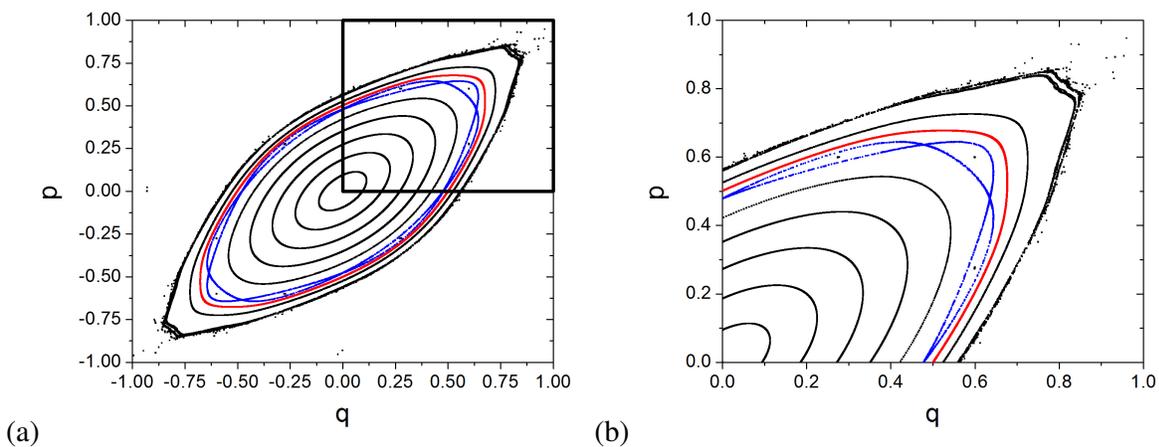

(a)  (b)

Figure 4 - (a) Phase space of the cubic map from equation (4) for the case $t = -1.1$. (b) Zoom in the green rectangle shown in (a). The blue curve indicates the separatrix of an island chain and the red curve a regular orbit.

Figure 4 shows the phase space of the cubic map from equation (4) for the case $t = -1.1$, which is an amplification of figure 3(d). The blue curve represents the separatrix of an island chain, and the red curve represents a regular orbit. Figure 4(b) shows a zoom in the rectangle in (a). One can notice chaotic orbits on the last surfaces.

## 4. Coupled cubic and quadratic map

In order to analyzed the chaotic motion around the invariant tori present near the origin of the cubic map, shown in figure 4, we will couple it with a two dimensional quadratic map through the generating function, $F = F_{cubic} + F_{quadratic} + F_{coupling}$, which is written as,

$$F(q, Q, q_{n+1}, Q_{n+1}) = QQ_{n+1} + qq_{n+1} + \frac{\tau Q^2}{2} - \frac{tq^2}{2} + \frac{Q^3}{3} + \frac{q^4}{4} + F_{coupling} \tag{5}$$

where $F_{coupling} = aqQ$ is the coupling function and the parameter $a$ defines its intensity. After calculating the new variables,

$$P = \frac{\partial F}{\partial Q}; \quad P_{n+1} = -\frac{\partial F}{\partial Q_{n+1}}; \quad p = \frac{\partial F}{\partial q}; \quad p_{n+1} = -\frac{\partial F}{\partial q_{n+1}} \tag{6}$$

we finally obtain the coupled cubic and quadratic map,

$$\begin{aligned} p_{n+1} &= -q \\ q_{n+1} &= p + qt - q^3 - aQ \\ P_{n+1} &= -Q \\ Q_{n+1} &= P - Q\tau - Q^2 - aq \end{aligned} \tag{7}$$

where $(q, p, Q, P)$ are the canonical variables of position and momentum. We choose the parameter $\tau = 0.9864$ as a small number, in order to avoid a strong influence of the quadratic map in the topology of the cubic map, and, by the same logic, we choose a weak coupling as well, $a = 0.01$.

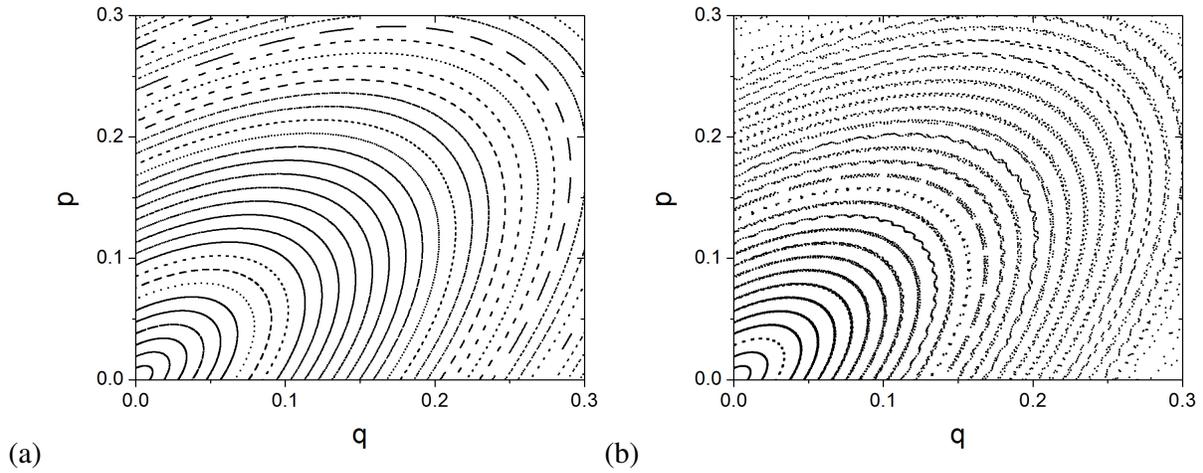

Figure 5 - Phase space of the map from equation (7) for the case $t = -1.1$ and $\tau = 0.9864$, at the plan $(q, p)$, where (a) Invariant tori for the uncoupled case, $a = 0$. (b) Invariant tori for the coupled case, $a = 0.01$.

Figure 5 shows the phase space of the map from equation (7) for the case $\tau = 0.9864$, at the plan $(q, p)$. We set $t = -1.1$ as an attempt to preserve the invariant tori around the elliptic fixed point at the origin, shown in figure 4, and, in order to emphasize the effect of the coupling on the topology of these tori, figure 5(a) shows the uncoupled case, $a = 0$, and figure 5(b) shows the scenario for a weak coupling case, $a = 0.01$. One can notice that the coupling affects the curves in such a way that they become "curly" when compared to the uncoupled case, fact that is expected, since the invariant tori, in the coupled case, are no longer two dimensional.

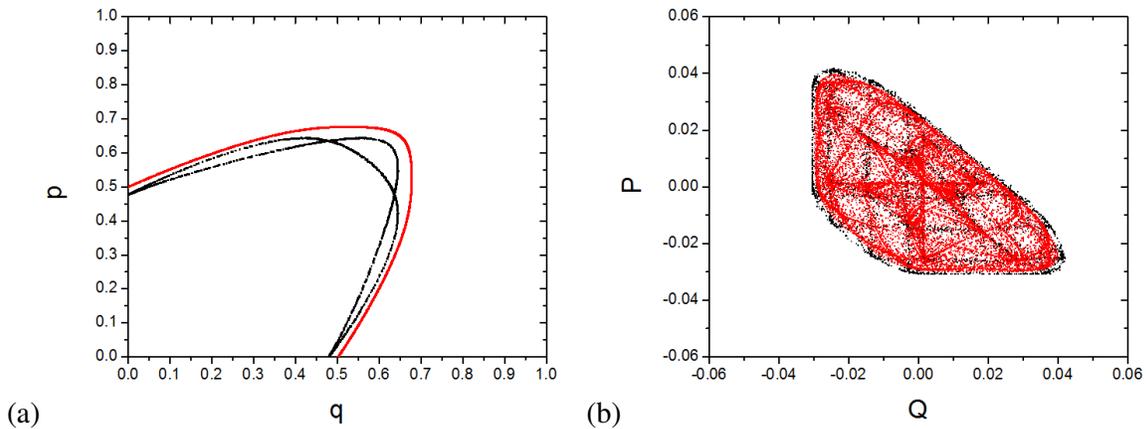

(a) (b)

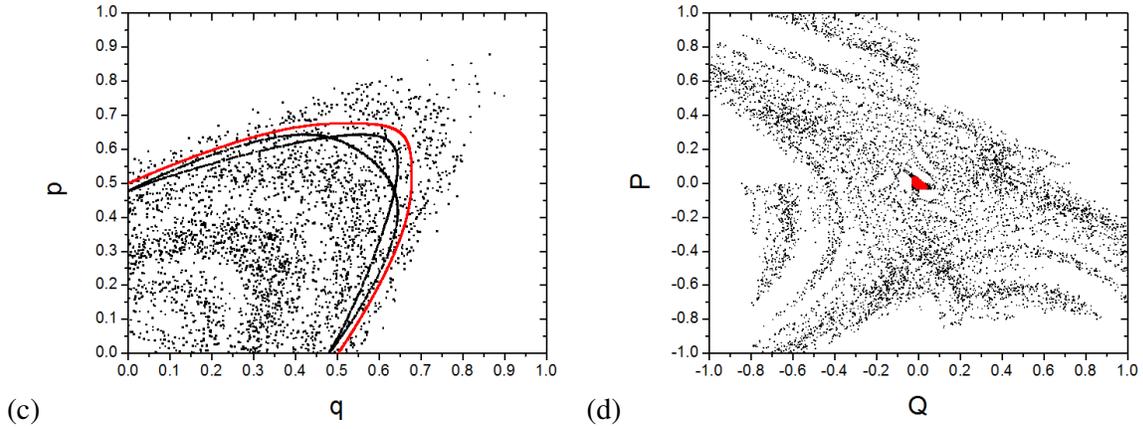

Figure 6 - Phase space of the coupled cubic and quadratic map from equation (7) for different plans and different numbers of iterations: (a) Phase space of the plan $(q,p)$ for $n = 5\times 10^3$ iterations. (b) Phase space of the plan $(Q,P)$ for $n = 5\times 10^3$ iterations. (c) Phase space of the plan $(q,p)$ for $n = 2\times 10^4$ iterations. (d) Phase space of the plan $(Q,P)$ for $n = 2\times 10^4$ iterations. The red curve represents a regular orbit and the black one represents a chaotic orbit.

Figure 6 shows the phase space of the coupled cubic and quadratic map from equation (7) with $\tau = 0.9864$, $t = -1.1$ and $a = 0.01$, for different numbers of iterations and different plans. Figures (a) and (b) were made for $n = 5\times 10^3$ iterations and figures (c) and (d) were made for $n = 2\times 10^4$ iterations. The regular orbit, in red, is the same regular orbit of the cubic map shown in figure 4, and, due to the weak coupling, the red orbit remains regular. The orbit in black is the same orbit related to the separatrix of the island chain in the cubic map of figure 4, shown in blue. We observe in figure (a) the regular orbit, in red, and the black orbit which lies on the separatriz of the island chain showed in figure 4, in blue. Figure (b) shows the same two orbits but in the plane $(Q,P)$. For a large number of iterations, $n = 2\times 10^4$, one can notice, in figure (c), that the orbit in red remains regular and does not change its topology, but the orbit in black reveals to be chaotic, and crosses the regular orbit in red. The same happens in figure (d), that shows the same orbits, but in the plan $(Q,P)$. Chaotic orbits do not cross regular orbits in two dimensional maps, but the same is not true in multidimensional maps, effect usually called as Arnold diffusion, then, the invariant tori surrounding the elliptic fixed point at the origin, that used to act as sub-neighborhoods in the stability proof, from page 2, no longer set the system as nonlinearly stable.

In order to verify the topology of the regular orbit, in red, from figure 6, we used the software GGobi to create a three dimension projection of our four dimensional object [24].

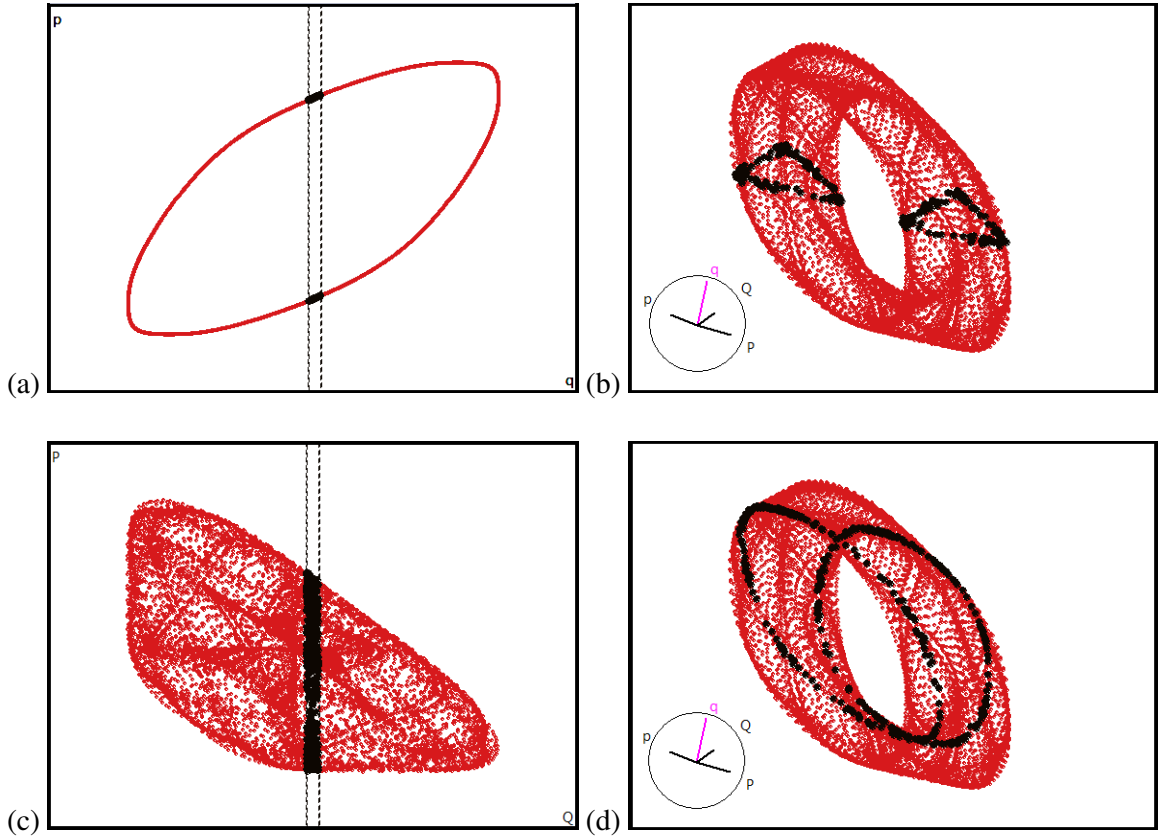

Figure 7 - GGobi simulations for the regular orbit, in red, of the coupled map from equation (7). (a) Regular orbit in the plane $(q, p)$. The dotted rectangle shows two small areas, in black, highlighted in (b). (b) Regular orbit in the three dimensional space, with a fixed $q$, in pink. The two black triangular areas are the same two black areas from figure (a), but in a different view. (c) Regular orbit in the plane $(Q, P)$. The dotted rectangle shows an area, in black, highlighted in (d). (d) The same regular orbit, but in the three dimensional space, with a fixed $q$, in pink. The two black elliptical areas in (d) are referent to the black area in (c), but in a different view.

Figure 7 shows GGobi simulations for the regular orbit, in red, from figure 6. Figure 7(a) shows the regular orbit in the plane $(q, p)$, which have an elliptical shape. The dotted rectangle shows two black areas, which are referent to the two triangular areas shown in 7(b). One can notice in 7(b) the same regular orbit, but in the three dimensional space, apparently represented by a torus. Figure 7(c) shows the regular orbit in the plane $(Q, P)$, where the black area covered by the dashed rectangle is emphasized in 7(d). One can observe in 7(d) the same regular orbit, but in the three dimensional space where the two black elliptical areas are referent to the black area in 7(c).

The four dimensional regular orbit, in red, when projected at the three dimensional space, has the topology of a triangular torus.

## 5. Conclusions

In two dimensional systems, invariant tori are closed curves which provide barriers to transport. If an elliptic fixed point is surrounded by a nested set of such curves, the equilibrium is nonlinearly stable. However, in higher dimensions, invariant tori do not provide barriers to the transport and allow an avenue for escape, which is usually called as Arnold diffusion, modifying the stability of the equilibrium. We analyzed this escape in the context of a four dimensional map, which was constructed by the coupling of a cubic and a quadratic map. The multidimensional map is polynomial, with a cubic degree of freedom that was designed to mimic the behavior of a Hamiltonian system with one positive and one negative energy mode, and a quadratic degree of freedom that allows eventual escapes. A special regular orbit present in the cubic degree of freedom is examined in detail when the coupling is activated; it was termed "triangular torus", and plays an important role in escape of chaotic trajectories.

**Acknowledgments**

The authors thank Dianne Cook and the other members of the GGobi group for useful discussions and comments. Caroline G. L. Martins thanks 'The National Council for Scientific and Technological Development' (CNPq) process number 211630/2013-6P. P. J. Morrison thanks the 'U.S. Department of Energy' Contract # DE-FG05-80ET-53088.